# Diseño de un controlador de ángulo en un Balancín.

**Alvarado Moreno, Jose David;**
**Delgadillo Romero, Kevin Andrey;**
**Galvis Reyna, David Enrique;**
**Poblador Parra, Gustavo Alonso;**
**Rodríguez Cortés, César Alejandro.**
*jalvarado@usbbog.edu.co*,
*kdelgadillo@academia.usbbog.edu.co*,
*degalvis@academia.usbbog.edu.co*,
*gpoblador@academia.usbbog.edu.co*,
*carodriguezc@academia.usbbog.edu.co.*

*Resumen* – **Este documento describe el diseño de un controlador PID para una planta tipo balancín de un grado de libertad. En el diseño de controlador se utilizarán los métodos de sintonización de Aström Hägglund (AH), Kaiser Chaira (KC) y Kaiser Rajka (KR) verificando el funcionamiento en simulaciones y en la planta. Por último, se describirá el desarrollo para la implementación de un controlador PID análogo mediante circuitos con amplificadores operacionales.**

*Palabras claves* — **Aström Hägglund (AH), Kaiser Chaira (KC), Kaiser Rajka (KR), control PID, Balancín, Matlab, Labview.**

# INTRODUCCIÓN

El controlador PID, como su nombre lo indica consta de tres etapas; Proporcional, Integral y Diferencial; este controlador es ampliamente utilizado en la industria, ya que se caracteriza por su simplicidad y robustez[1], sus múltiples configuraciones y lo más importante que puede soportar perturbaciones en el sistema y controlar el actuador para que la planta vuelva a el punto de referencia establecido o set-point.

Actualmente diferentes investigaciones se han desarrollado para verificar las técnicas de sintonización del controlador PID sobre una planta como el balancín, la Universidad de Valladolid, muestra el diseño de un controlador asociado a dicha planta, se verifico el funcionamiento del controlador junto con las simulaciones realizadas [2][3]. En Colombia se han realizado investigaciones para controlar la posición del balancín mediante tarjetas embebidas en dichos estudios se muestra que la simulación corresponde con el comportamiento real de la planta [4][5].

En este artículo se explicará cómo se diseña un controlador, obteniendo la funcione de transferencia de la planta manera experimental, para realizar por diferentes métodos la sintonización del controlador y diseñar el circuito de manera analógica para un balancín.

## DISEÑO DEL EXPERIMENTO

El diseño del controlador depende de la respuesta dinámica de la planta, en este caso se utilizará un balancín que será la planta a la cual se le desarrollará el sistema de control. El diseño de plata se ilustra en la Figura 1, la cual está compuesta por un motor DC a 12 v que será el encargado de realizar el movimiento de la hélice, un potenciómetro lineal de 10K Ω que se utilizara como sensor de Angulo, y para la construcción se utilizó varillas de acero inoxidable, ángulos para soporte y una chumacera que servirá como eje de rotación.

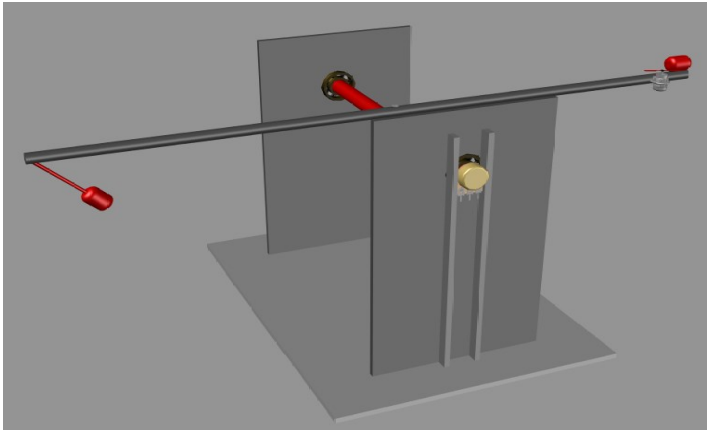

***Fig. 1 -*** *Dibujo de la planta en AutoCAD. Fuente: Autor*

En la Figura 2 se observa el balancín construido tiene una altura de 42.5 cm, por 40.5 de ancho. Dicho sistema por controlar está compuesto por la estructura básica en triángulo, una chumacera, dos varillas, una de 30 cm y 90 cm, un potenciómetro de 10 KΩ, un motor de 12 V-DC y dos contrapesos ubicados en los extremos, cada uno de 40 gramos.

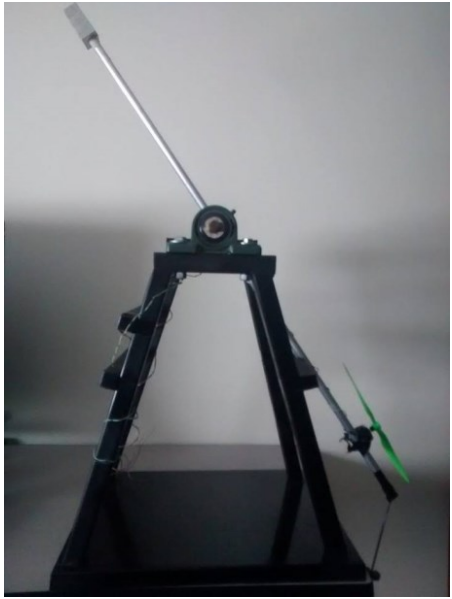

***Fig. 2 -*** *Planta Balancín. Fuente: Autor*

Luego de construir la planta se procede a realizar el modelo y el sistema de control, a continuación, se describe la obtención del modelo en forma experimental y el diseño del controlador para el balancín construido.

## IDENTIFICACIÓN DEL MODELO DINÁMICO DE LA PLANTA

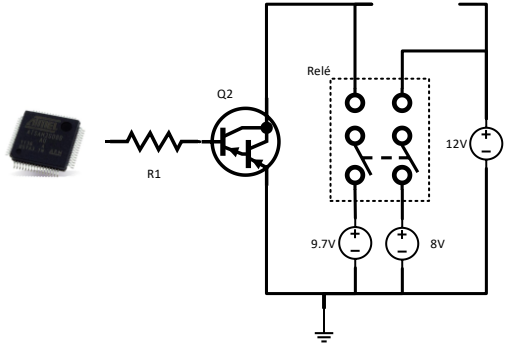

*Fig. 3 -* Esquema del circuito. Fuente: Autor.

Para la identificación de la planta es necesario programar un microcontrolador ATMEGA16A con la finalidad de generar los pulsos en el actuador, seguido de un Relay que proporciona los valores mínimos (8v) y máximos (9.7v) del voltaje del pulso. El esquema del circuito generador de pulsos está compuesto por un relay que proporciona los dos valores de voltajes en el momento del estado del pulso y un transistor NPN que regula la potencia de salida, como se ilustra en la Figura 3.

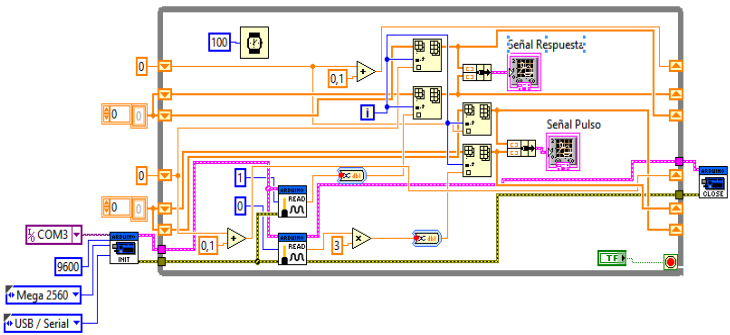

*Fig. 4 -* Diagrama de adquisición de Datos LabVIEW. Fuente: Autor.

65

La adquisición de datos se realizó mediante la tarjeta Arduino en LabVIEW, configurando las entradas análogas, para obtener la señal de pulso y la señal de respuesta de la planta en tiempo real por medio de conversión por arreglos como se ve en la Figura 4. Para obtener los datos que registra la tarjeta, en la Figura 5 se muestran los datos graficados.

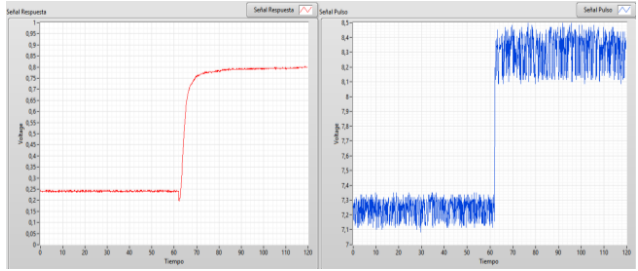

**Fig. 5 -** Captura de Datos en LabVIEW. Fuente: Autor.

Con los datos obtenidos en LabVIEW se almacenan en un archivo Excel y se exportan a Matlab para calcular la función de trasferencia. La Figura 6 corresponde al primer registro de datos, en donde la parte izquierda es la respuesta de la planta y la derecha corresponde al pulso de entrada.

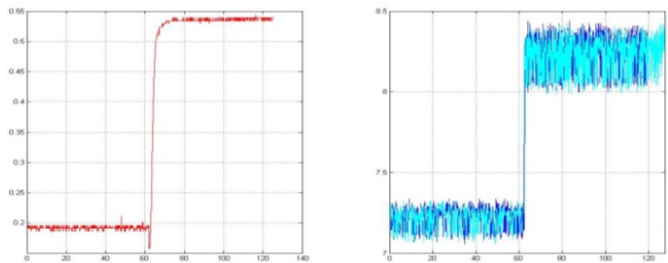

**Fig. 6 -** Respuesta al pulso. Fuente: Autor.

En la Figura 7 se grafican las dos últimas pruebas, en donde se observa que la planta se estabiliza alrededor de los 60 segundos con una ganancia de alrededor de 0.48.

Con cada una de las tomas se realiza un promedio para determinar la ecuación de trasferencia de la planta. En la Figura 8 se grafican las diferentes señales de entrada (pulso) en una gráfica para realizar el promedio.

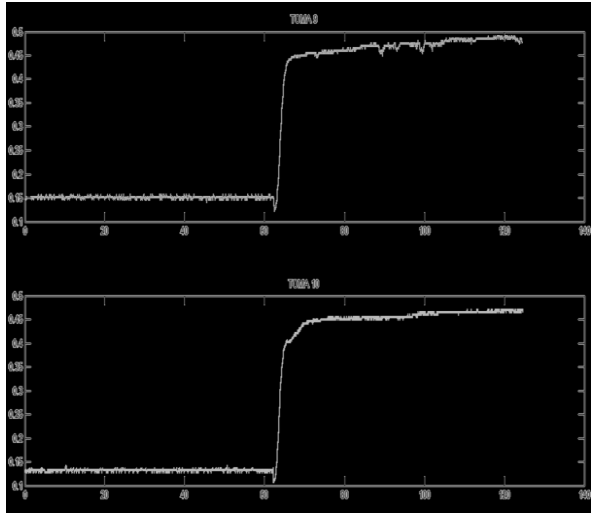

*Fig. 7 -* Señal de repuesta de la planta. Fuente: Autor.

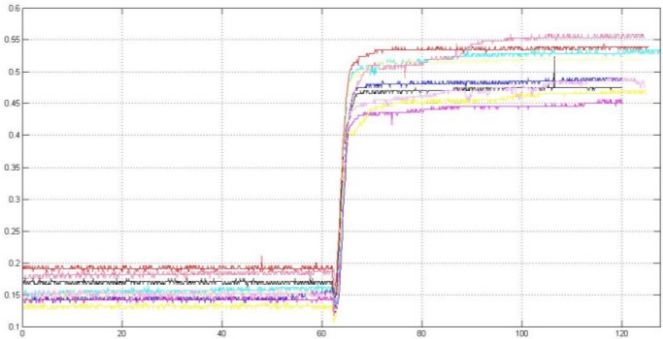

*Fig. 8 -* Señales de respuesta de la planta para realizar el promedio. Fuente: Autor.

En la Figura 8 se grafican las 10 respuestas obtenidas y en la Figura 9 se grafica la señal de respuesta promedio obtenida.

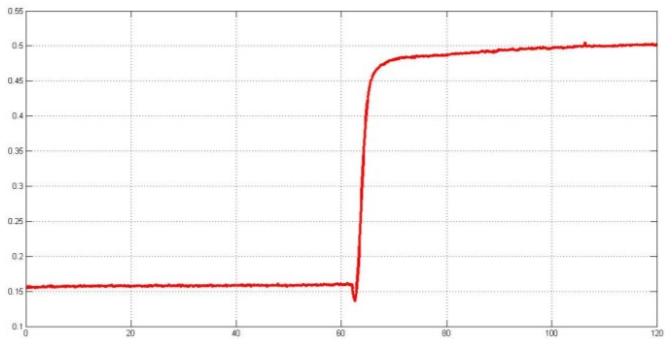

*Fig. 9 -* Señal de promedio de la respuesta. Fuente: Autor.

Para caracterizar la función de transferencia de la planta existen métodos como; determinístico, estocástico o basado en la respuesta paso[6], como los propuestos por Van Der Grinnten [7] o el método de Smith[8], este último fue el utilizado debido a que era adecuado para el tipo de respuesta del sistema, y que consiste en trazar una recta tangente a la respuesta y medir los puntos en donde dicha recta corta la gráfica como se muestra en la Figura 9.

A partir de la respuesta promedio de la planta se obtienen los valores del tiempo de respuesta (Tao) (ecuación 3), la ganancia (ecuación 1) y el tiempo de retraso (ecuación 2)[9].

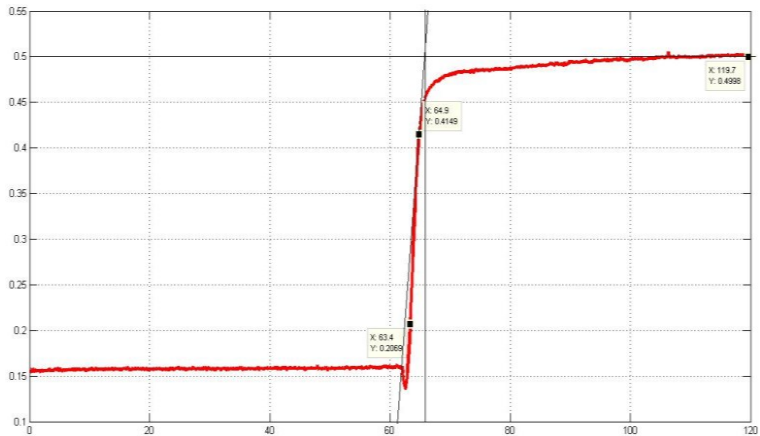

*FIG. 10* - Señal de promedio de la respuesta. Fuente: Autor.

$$K_P = \frac{0.4895 - 0.1569}{1.033} = 0.322 \quad (\mathbf{1})$$

$$\tau_d = 63.43 - 64.76 = 1.3 \quad (\mathbf{2})$$

$$\tau = 62.9 - 61.6 = 1.33 \quad (\mathbf{3})$$

Una vez obtenidos estos valores, se remplazarán para obtener la función de trasferencia de la planta de la siguiente manera.

$$G(s) = \frac{K_P}{\tau s + 1} e^{-\tau_d s} \quad \textbf{(4)}$$

$$G(s) = \frac{0.322}{1.33s + 1} e^{-1.3s} \quad \textbf{(5)}$$

En la Figura 11 se muestra la comparación de las gráficas creadas por medio de las ecuaciones y las de la planta. Se observa en la parte superior izquierda la señal de entrada escalón creada y a la derecha la obtenida. En la parte inferior se compara la señal de respuesta creada y obtenida.

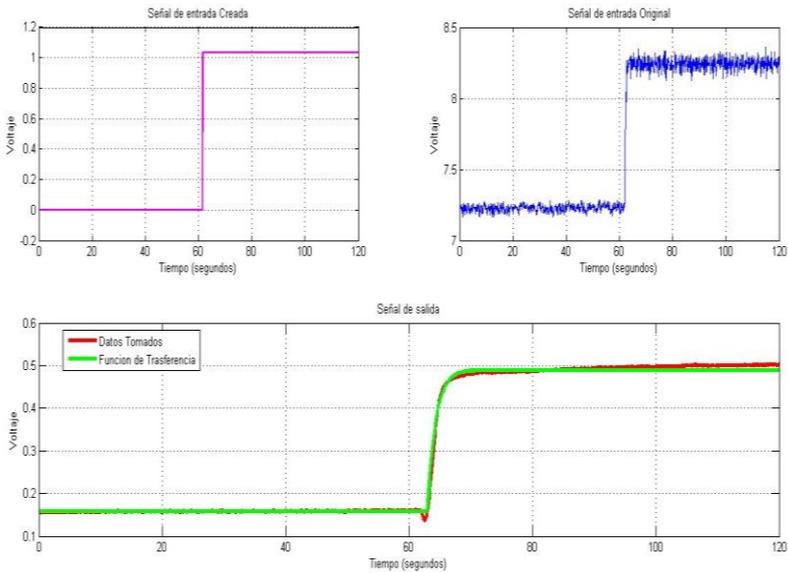

*Fig. 11* - Señal de Datos tomados y función de transferencia comparadas. Fuente: Autor.

Una vez obtenido el modelo de la planta se procede a diseñar el controlador. En la Figura 12 se ilustra el diagrama de control en lazo cerrado[10] que se utilizó. En donde el sensor será el potenciómetro lineal, el actuador será un puente H y el controlador un PID.

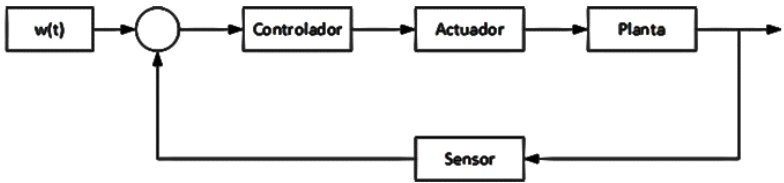

**FIG. 12** - *Diagrama de bloques Control en lazo cerrado. Fuente: Autor.*

## DISEÑO DEL SISTEMA DE CONTROL PID

El diseño del controlador se realizó utilizando el método de la realimentación con relé, el en donde se busca que el proceso oscile permanente par cual ha sido empleado eficientemente para sintonizar controladores PID [11]. A continuación, se describe el procedimiento de diseño para estos métodos[12].

A. **Método De Aströn Hägglund (AH)**

El modelo consiste en conectar en un lazo realimentado con un relé como se muestra en la Figura 13. En donde se busca que sistema oscile continuamente como se muestra en la Figura 14, la señal de salida obtenida se aproxima una señal sinodal [13].

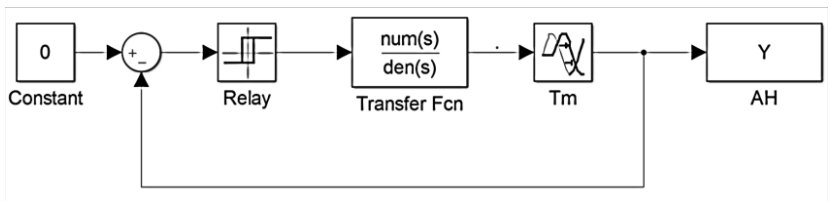

**FIG. 13** - *Función de trasferencia de la planta en realimentación con relé. Fuente: [11].*

La amplitud y el periodo que se produce en la señal de permiten diseñar establecer las ganancias de un controlador PID a partir de unos criterios de diseño definidos [14].

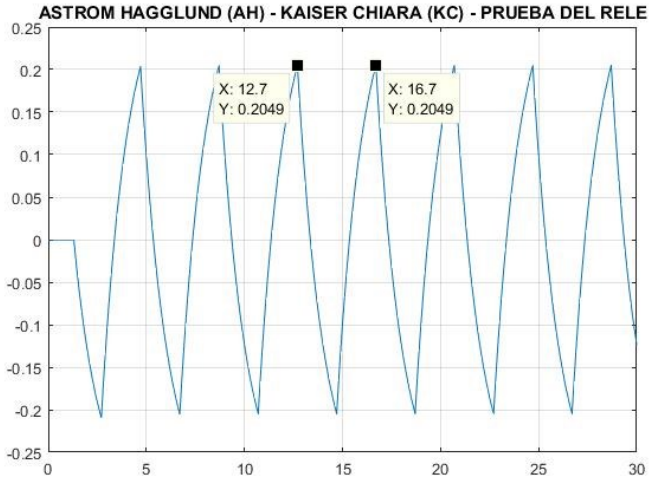

***FIG. 14*** - *Salida de la planta en realimentación con relé. Fuente: Autor.*

A partir de la respuesta de oscilación de la figura 14, se realiza la medición del periodo crítico (***Tc***) y la amplitud (***a***), para un valor de conmutación igual a uno (***d=1***) se definen en las ecuaciones 6 y 7.

$$a = 0.2049 \quad (6)$$

$$T_C = 16.7 - 12.7 = 4.0 \quad (7)$$

Con el valor de la amplitud se obtiene la ganancia crítica como se ve a continuación en la ecuación 9[15]

$$K_C = \frac{4d}{\pi a} = \frac{4(1)}{\pi(0.2049)} = 6.21 \quad (8)$$

Al obtener el valor de la ganancia crítica (*Kc*) y del periodo crítico (*Tc*) se pueden calcular las ganancias del controlador ganancia proporcional, integral, derivativa del controlador PID se calculan a partir de las ecuaciones 10, 11 y 12[16].

$$K_P = 0.6 * K_C = 0.6 * 6.21 = 3.72 \quad (9)$$

$$T_i = 0.5 * T_C = 0.5 * 4.0 = 2.0 \quad (\mathbf{10})$$

$$T_d = 0.25 * T_i = 0.25 * 2.0 = 0.5 \quad (\mathbf{11})$$

Una vez obtenidas las ganancias se procede a simular el controlador PID utilizando la estructura no interactuante[11], como se muestra en la Figura 15.

En la simulación se definió una perturbación a partir de una señal paso a los 50 segundos, el resultado de la simulación se observa en la Figura 16, donde el controlador se estabiliza en 25s y en la perturbación requiere 20s para retomar el set-point y un sobreimpulso del 40%.

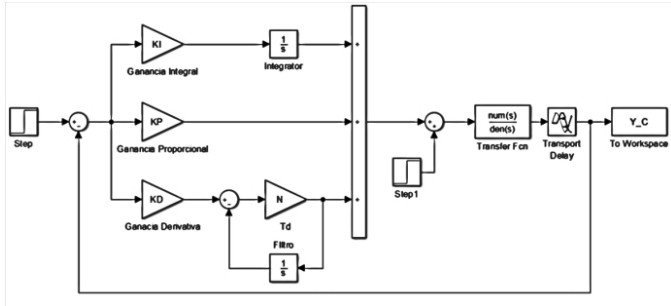

*Fig. 15* - Diagrama de bloques controlador PID no interactuante. Fuente: Autor.

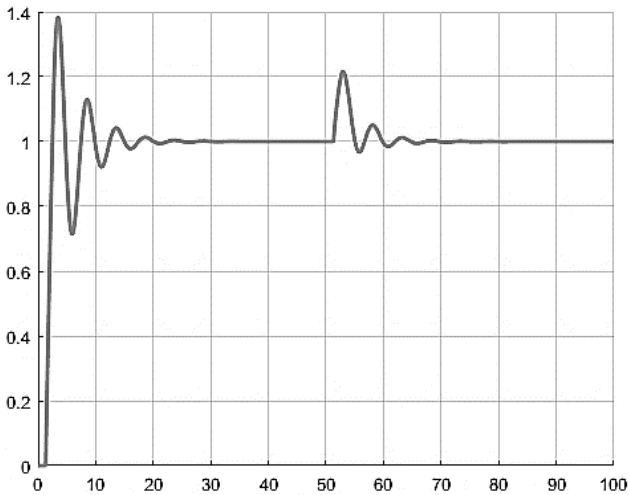

*Fig. 16* - Respuesta del Control PID sintonizado con el método Aströn Hägglund. Fuente: Autor.

## B. Método De Kaiser Chiara (KC).

En el método de **KC** [17] se realiza la prueba de realimentación de relé como en **AH**, por lo que el valor de periodo crítico (*Tc*), la amplitud (*a*) y el valor de conmutación (d) corresponde a los de las ecuaciones 6, 7 y 8. Para este método se complementa con el criterio de diseño adicional denominado margen de fase (***PM***), el cual en definido en un rango de 40° y 70° [13], en este trabajo se definió de 50°.

El valor de la ganancia (*Kc*) y periodo (*Tc*) crítico serán iguales a los de **AH,** las ganancias proporcional integral y derivativa del controlador PID se calculan mediante las ecuaciones 12, 13, 14.

$$K_P = K_C \cos(P_M) = 6.21 * \cos(50) = 3.99 \quad (\mathbf{12})$$

$$T_i = T_C \frac{1 + sen(P_M)}{\pi \cos(P_M)} = 4.0 \frac{1 + sen(50)}{\pi \cos(50)} = 3.49 \quad (\mathbf{13})$$

$$T_d = 0.25 * T_i = 0.25 * 3.07 = 0.8745 \quad (\mathbf{14})$$

Se realiza la simulación del controlador utilizando la estructura del modelo de la Figura 15.

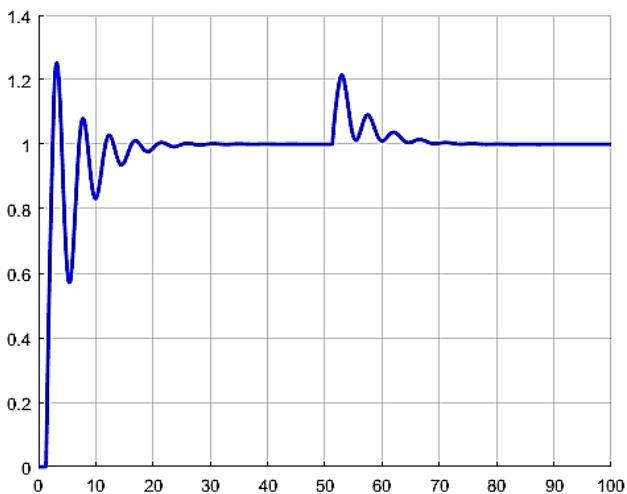

*Fig. 17* - Respuesta del Control PID sintonizado con el método Kaiser Chiara. Fuente: Autor.

73

El resultado de la simulación se observa en la Figura 17, en donde el controlador se estabiliza en 30s y en la perturbación requiere 20s para retomar el set-point y un sobre-impulso del 22%.

### C. Método De Kaiser Rajka (KR)

El método de KC [15] se compone de dos partes, la primera es efectuar la prueba de realimentación de relé de **AH** y la segunda es realizar nuevamente la prueba de realimentación de relé con un **Transport Delay** ($Td$) adicional, como se muestra en la Figura 18.

Para establecer el valor $\tau_d$ se calcula a partir del valor del perdido critico ($Tc = 3.2$) obtenido en prueba AH y de un parámetro de diseño de margen de fase *($P_M$ = 70)*. El cálculo se realiza a partir de la ecuación 15.

$$\tau_d = \frac{P_M - 37}{360} T_C = \frac{70 - 37}{360}(4.0) = 0.36 \quad (15)$$

Con el valor del tiempo muerto (Td) se configura en el bloque **Transport Delay** de la Figura 18.

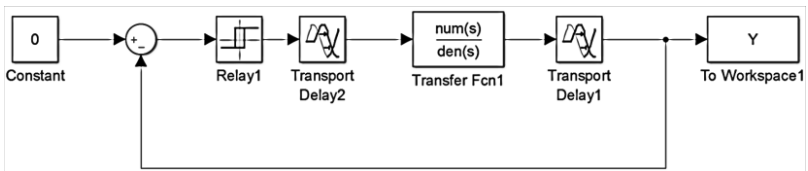

*Fig. 18* - *Planta en realimentación con relé con tiempo muerto. Fuente:[13].*

Al realizar la simulación de la prueba del relé se obtiene la respuesta de oscilación de la Figura 19, en donde se realiza la medición del periodo crítico (***Tc***) y la amplitud (***a***), para un valor de conmutación igual a uno (***d=1***), como se define en las ecuaciones 16 y 17.

$$a = 0.2272 \quad (16)$$

$$T_C = 15.3 - 10.5 = 4.8 \qquad (\mathbf{17})$$

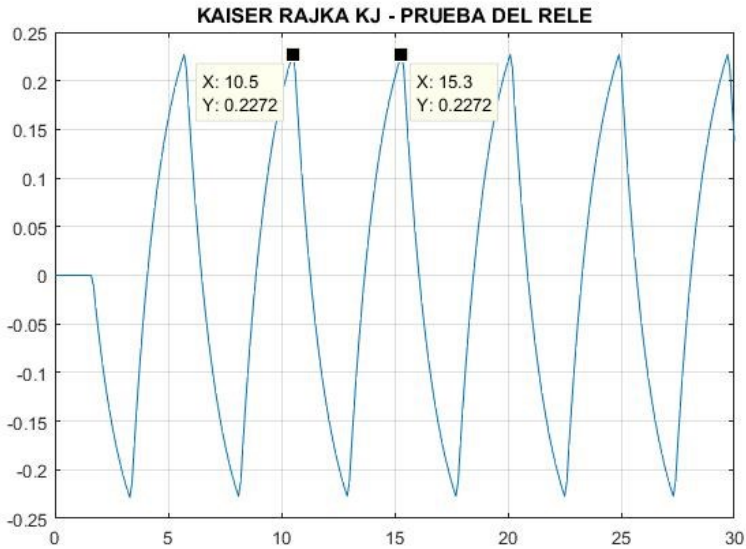

*FIG. 19* - *Salida de la planta en realimentación con relé con tiempo muerto. Fuente: Autor.*

Con el valor de la amplitud se obtiene la ganancia crítica como se ve a continuación:

$$K_C = \frac{4d}{\pi a} = \frac{4(1)}{\pi(0.2272)} = 5.6 \quad (\mathbf{18})$$

Al obtener el valor de la ganancia crítica (*Kc*) y del periodo crítico (*Tc*) se pueden calcular las ganancias del controlador PID mediante las ecuaciones 19, 20, 21[13].

$$K_P = 0.8 * K_C = 0.8 * 5.6 = 4.48 \quad (\mathbf{19})$$

$$T_i = 0.64 * T_C = 0.64 * 4.8 = 3.072 \quad (\mathbf{20})$$

$$T_d = 0.25 * T_i = 0.25 * 3.072 = 0.768 \quad (\mathbf{21})$$

Se realiza la simulación del controlador utilizando la estructura del modelo de la Figura 15, los resultados obtenidos se observan en la Figura 20, donde el controlador se estabiliza en 35s y en la perturbación requiere 30s para retomar el set-point y un sobreimpulso del 42%.

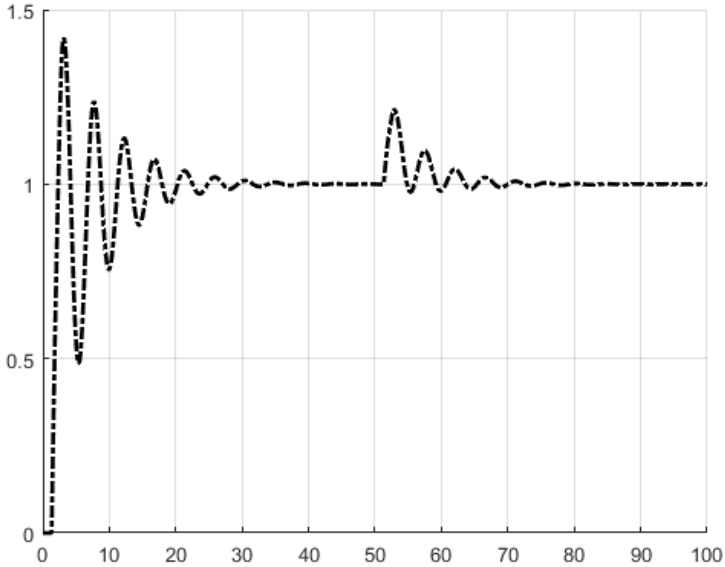

F<small>IG</small>. *20 - Respuesta del Control PID sintonizado con el método Kaiser Rajka. Fuente: Autor.*

Se realiza una comparativa entre cada una de las respuestas como se ve en la Figura 21, con el objetivo de escoger cual método presento mejores resultados.

## I<small>MPLEMENTACIÓN DEL CONTROLADOR</small>.

El controlador este compuesto por un circuito que realiza la comparación del set-point y la señal de salida del sensor, en donde se obtiene la señal de error que será la entrada del controlador. Para implementar el controlador se requiere implementar un circuito adicional en donde se incorporarán las ganancias del controlador diseñado.

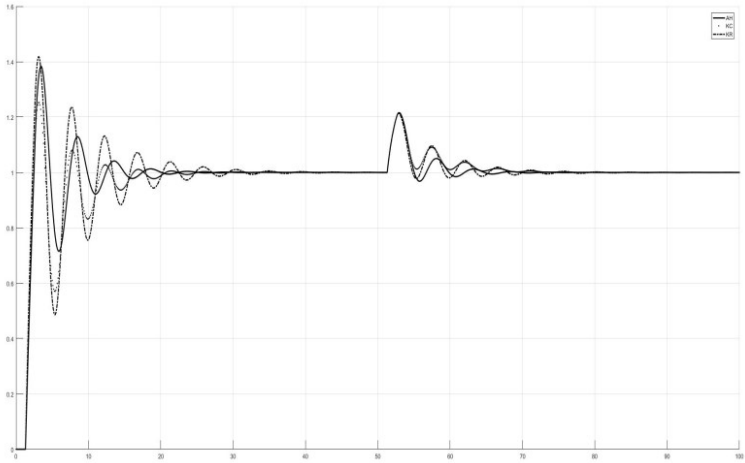

*Fig. 21* - *Comparativa de los métodos de sintonización PID. Fuente: Autor.*

A. **Circuito comparador**

Para conseguir la señal de error se empleó un circuito restador con amplificadores operacionales, los valores de las resistencias deben ser iguales para que se realice la resta en este caso se emplearon resistencias de 470 Ω. En la Figura 22 se muestra el diagrama del circuito.

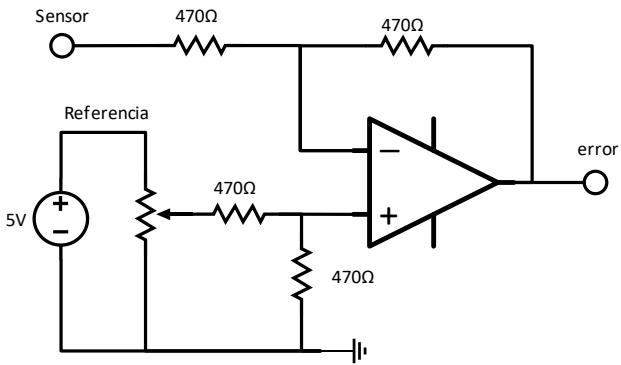

*Fig. 22* - *Circuito Restador. Fuente: Autor.*

B. **Circuito controlador PID**

Se implementara el circuito de la Figura 23[10] para el controlador PID análogo, en el cual las ganancias esta definidas en las ecuaciones 22, 23 y 24.

77

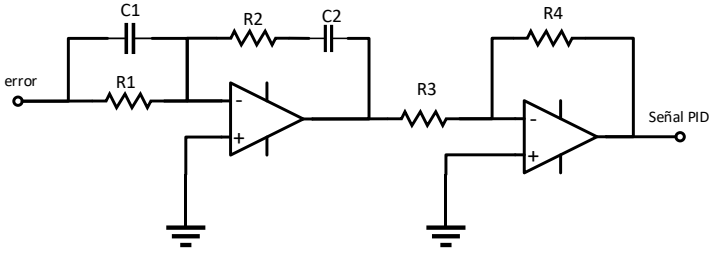

**FIG. 23** - *Circuito Restador. Fuente: Autor.*

$$KP = \frac{R_4(R_1C_1 + R_2C_2)}{R_3R_1C_2} \quad (22)$$

$$KI = \frac{R_4}{R_3R_1C_2} \quad (23)$$

$$KD = \frac{R_4R_2C_1}{R_3R_1C_2} \quad (24)$$

Debido a que las ganancias del controlador se definen en términos de KP, KI y KD, es necesario definir estos parametros a partir de las constantes, como se muestra en las ecuaciones 25, 26 y 27.

$$KP = K_{PKC} = 3.99 \quad (25)$$

$$KI = \frac{K_{PKC}}{T_{iKC}} = \frac{3.99}{2.04} = 1.95 \quad (26)$$

$$KD = K_{PKC} * T_{dKC} = 3.99 * 0.51 = 2.03 \quad (27)$$

Para calcular los valores se procede a despejar los valores de las resistencias R1 y R2, para lo que se establecen los valores de los condensadores $C_1 = 220\mu F$ y $C_2 = 22\mu F$.

El valor de $R_1$ y $R_2$ se obtiene a partir de la ecuación 28 y 29.

$$R_1 = \frac{KP}{2KIC_2} = \frac{3.99}{(2)(1.95)(22\mu F)} \cong 45K\Omega \qquad (\mathbf{28})$$

$$R_2 = \frac{KP}{2KIC_1} = \frac{3.99}{(2)(1.95)(220\mu F)} \cong 4.5K\Omega \qquad (\mathbf{29})$$

Para obtener $R_4$ se define $R_3 = 1K\Omega$ y se despeja $R_4$ en la ecuación 30.

$$R_4 = KIR_3R_1C_2 = (1.95)(1K\Omega)(45K\Omega)(22\mu F) \cong 1.9K\Omega \quad (\mathbf{30})$$

### C. Integrador no inversor con alto Factor de calidad

Adicionalmente se agregó un integrador no inversor con alto factor de calidad con valores para $R_1$ de 220 Ω y $R_2$ de 550KΩ además de un condensador de 100pF.

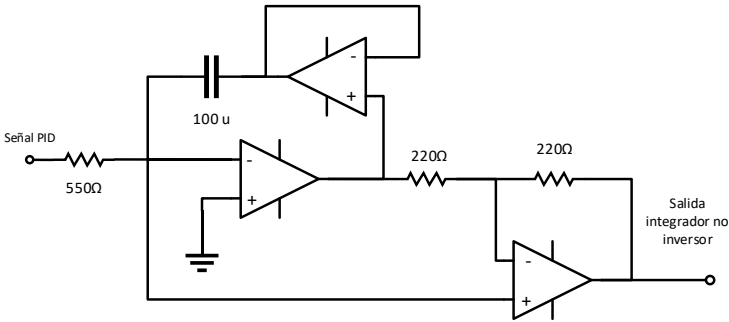

***FIG. 24*** - *Integrador no inversor con alto factor de calidad. Fuente: ASC[1].*

### D. Actuador

El actuador recibe la señal del controlador y con ella hace que la planta actúe, en este caso, se empleó un PWM de 8 bits sobre un microcontrolador ATmega 16 y junto a este un puente H, el cual suministrará la corriente y el voltaje necesarios para que el motor pueda elevarse al punto deseado.

---

[1] Análisis y síntesis de circuitos, departamento de electrónica y electromagnetismo, universidad de Sevilla, Apéndice 2, Bloques Básicos Activos.

Para obtener los resultados se unen todos los circuitos cada uno entregando las señales y voltajes correspondientes para un correcto funcionamiento. Finalmente se sintonizo el circuito en físico con las resistencias obtenidas y seleccionadas se comprobó el funcionamiento del controlador.

# CONCLUSIONES.

El método que presento mejor desempeño en términos de tiempo de establecimiento sobre-impulso es Kaiser Chiara. Además, la robustez del controlador en termino de respuesta ante una perturbación es la que necesita menor tiempo para regresar a el punto de operación.

Es necesario emplear un integrador no inversor con alto factor de calidad debido a que un modelo integrador con operaciones común no funcionaba adecuadamente y generaba oscilaciones no controladas. La planta actuó de acuerdo a la simulación realizada, presentaba una respuesta y tiempo de estabilización similar ante una perturbación.

## REFERENCIAS

[1]. J.-C. Shen, "New tuning method for PID controller," ISA Trans., vol. 41, no. 4, pp. 473–484, Oct. 2002.

[2]. V. Viltres La Rosa and F. J. García Ruiz, "Control de posición de un balancín con motor y hélice," 2012.

[3]. Á. Martín Ballesteros and M. del Río Carbajo, "Control de posición de un balancín con Arduino," 2013.

[4]. E. Alejandro, P. Castiblanco, and P. Edwin, "Control de posición de un balancín Motor-Hélice," no. August 2015, pp. 0–14, 2016.

[5]. A. F. Caro, F. J. López, and J. J. Ortíz, "Control Multivariable para un Helicóptero de dos grados de libertad utilizando Algoritmos Genéticos de Control," p. 160, 2005.

[6]. J. Mikleš and M. Fikar, "Process Identification," in Process

Modelling, Identification, and Control, Berlin, Heidelberg: Springer Berlin Heidelberg, pp. 221–251.

[7]. S. Martinez, H. Numpaque, and J. Alvarado, "Efecto de la Temperatura en la Producción de Biogás en un Bioreactor tipo Batch a través de la Descomposición Anaeróbica de Residuos Sólidos Orgánicos," ENGI Rev. Electrónica la Fac. Ing., vol. 3, no. 1, pp. 16–19, 2016.

[8]. I. Dale E. Seborg, Thomas F. Edgar, Duncan A. Mellichamp, Francis J. Doyle, Process dynamics and control. John Wiley & Sons, Inc, 2016.

[9]. B. Criollo, J. Alvarado, and H. Numpaque, "PID temperature control and pH dosage for methane produccion from anaerobic digestion of organic solid waste," Rev. Colomb. Tecnol. Av., vol. 2, no. 26, pp. 134–141, 2014.

[10]. K. Ogata, Ingeniería de control moderna. Pearson Educación, 2010.

[11]. K. J. Åström and T. Hägglund, Control PID avanzado. Pearson Prentice Hall, 2009.

[12]. J. D. Alvarado Moreno, "Diseño de un sistema de control de temperatura y pH, en el proceso de digestión anaeróbica para residuos sólidos orgánicos en un bio-reactor tipo BATCH." Universidad de Ibague, 2015.

[13]. R. De Keyser and C. Ionescu, "A Comparative Study of Three Relay-based PID Auto-Tuners," Model. Identification, Control, no. AsiaMIC, 2010.

[14]. Q.-G. Wang, B. Zou, T.-H. Lee, and Q. Bi, "Auto-tuning of multivariable PID controllers from decentralized relay feedback," Automatica, vol. 33, no. 3, pp. 319–330, 1997.

[15]. I. A. R. Ruge, "Optimización de señal de control en reguladores PID con arquitectura antireset Wind-Up," no. 30, pp. 24–31, 2011.

[16]. J. Nandong, "Heuristic-based multi-scale control procedure of simultaneous multi-loop PID tuning for multivariable processes," J. Process Control, vol. 35, pp. 101–112, 2015.

[17]. V. M. Alfaro Ruiz, "Método de sintonización de controladores que operan como servomecanismos," Ing. Rev. la Univ. Costa Rica, vol. 13, pp. 13–29, 2003.